\documentclass[10pt,twocolumn,prl,superscriptaddress,showpacs,floatfix,aps]{revtex4-2}

\usepackage{comment}

\usepackage{lmodern}
\usepackage[colorlinks=true, allcolors=blue]{hyperref}
\usepackage{graphicx}
\usepackage{dcolumn}
\usepackage{overpic}
\usepackage{amsmath,amssymb,amsmath,bm,mathtools}
\usepackage[svgnames]{xcolor}

\newcommand{\bprop}{F}
\newcommand{\de}{\mathrm{d}}

\newcommand{\mycomment}[1]{}

\begin{document}

\title{Learning Backward Transport for Source Localization}

\author{Maurizio Carbone}\email{maurizio.carbone.2@uniroma2.it}
\affiliation{Istituto dei Sistemi Complessi, CNR, Via dei Taurini 19, 00185 Rome, Italy}
\affiliation{Department of Physics \& INFN, Tor Vergata University of Rome, Via della Ricerca Scientifica 1, 00133 Rome, Italy}
\author{Lorenzo Piro}
\affiliation{Department of Physics \& INFN, Tor Vergata University of Rome, Via della Ricerca Scientifica 1, 00133 Rome, Italy}
\affiliation{Department of Technological Innovations and Safety of Plants, Products and Anthropic Settlements (DIT), Italian National Institute for Insurance against Accidents at Work, INAIL, Rome, Italy}

\date{\today}
 
\begin{abstract}
    We address the problem of locating a chemical source in a flow. Based on the duality between the concentration field and Lagrangian tracer trajectories, we interpret concentration detections as evidence of paths connecting the source to the detection points. This Schrödinger bridge formulation between plausible emission positions and detection points leverages the backward propagator of passive tracers to frame source localization as the sampling of candidate emission locations via Langevin dynamics. The associated drift reveals classical chemotaxis and cast-and-surge as complementary behaviors emerging from a single transport-based principle. Applied to olfactory search in two-dimensional turbulence, the proposed backtracking framework outperforms classical strategies across varying wind regimes using a single, Galilean-invariant learned propagator.
\end{abstract}

\maketitle

Locating the source of a chemical transported by fluid flows from local sparse detections is a fundamental inverse problem in physics~\cite{Berg1993,Celani2014}, with applications ranging from biological olfaction \cite{Murlis1992,Hansson2011,Haverkamp2018,Carde2021,Reddy2022} to environmental monitoring and autonomous search-and-rescue \cite{Burgues2020,Jing2021,Mansfield2024}.
A crucial physical aspect of this challenge lies in the Lagrangian description of passive scalar transport: the Eulerian chemical concentration field at any given point is determined by the trajectories of passive tracers evolving backward in time to the source \cite{Kac1949,Falkovic2001,Celani2004}. 
Despite this underlying physical duality, traditional search strategies focused on optimizing heuristic or probabilistic policies rather than modeling the transport process itself.
Classical bio-inspired heuristics, such as chemotaxis~\cite{Berg1972,Celani2010} and cast-and-surge~\cite{Balkovsky2002}, prescribe responses based on local chemical concentration, but rely on flow conditions under which gradients or mean wind provide reliable guidance. Information-theoretic approaches, such as Infotaxis~\cite{Vergassola2007,Loisy2022,Heinonen2025}, formulate source localization as a partially observable Markov decision process~\cite{Kaelbling1998,Heinonen2022,Carbone2026}, selecting actions that maximize the expected information gain at the cost of maintaining and updating a probability map over the entire search domain.
More recently, finite-state controllers~\cite{Verano2023} and reinforcement learning~\cite{Hartl2021,Singh2023,Rando2025,Piro2026} have demonstrated that effective search strategies can be learned directly from iterative interaction with the environment. However, these approaches optimize policies for specific conditions, limiting transferability across flow regimes and, especially for reinforcement learning, often sacrificing physical interpretability.

We take a different perspective: instead of learning a search strategy, we learn the transport process itself. During an offline phase, the agent infers the backward propagator of passive tracers---the conditional probability that a detected particle originated from a given location---directly from observations of tracer transport.
Source localization is then recast as an inference problem: each detection provides evidence for tracer trajectories connecting the detection point to the source, and localization follows from sampling the resulting distribution of candidate emission locations, rather than from an explicitly optimized strategy.
This viewpoint naturally connects source localization to Schr\"odinger bridge formulations of stochastic transport~\citep{Schroedinger1931,Villani2008}.
Upon each detection, the agent uses the learned propagator to construct the distribution of plausible emission points and samples it via Langevin dynamics. The corresponding drift is obtained by solving an inverse Fokker--Planck problem.
Classical behaviors emerge naturally within this framework: surge follows the evolution of the propagator mean, casting arises from divergence-free probability currents exploring propagator contour levels, and chemotaxis corresponds to ascent along its gradient.

 \begin{figure*}
    \centering
    \includegraphics[width=.96\textwidth]{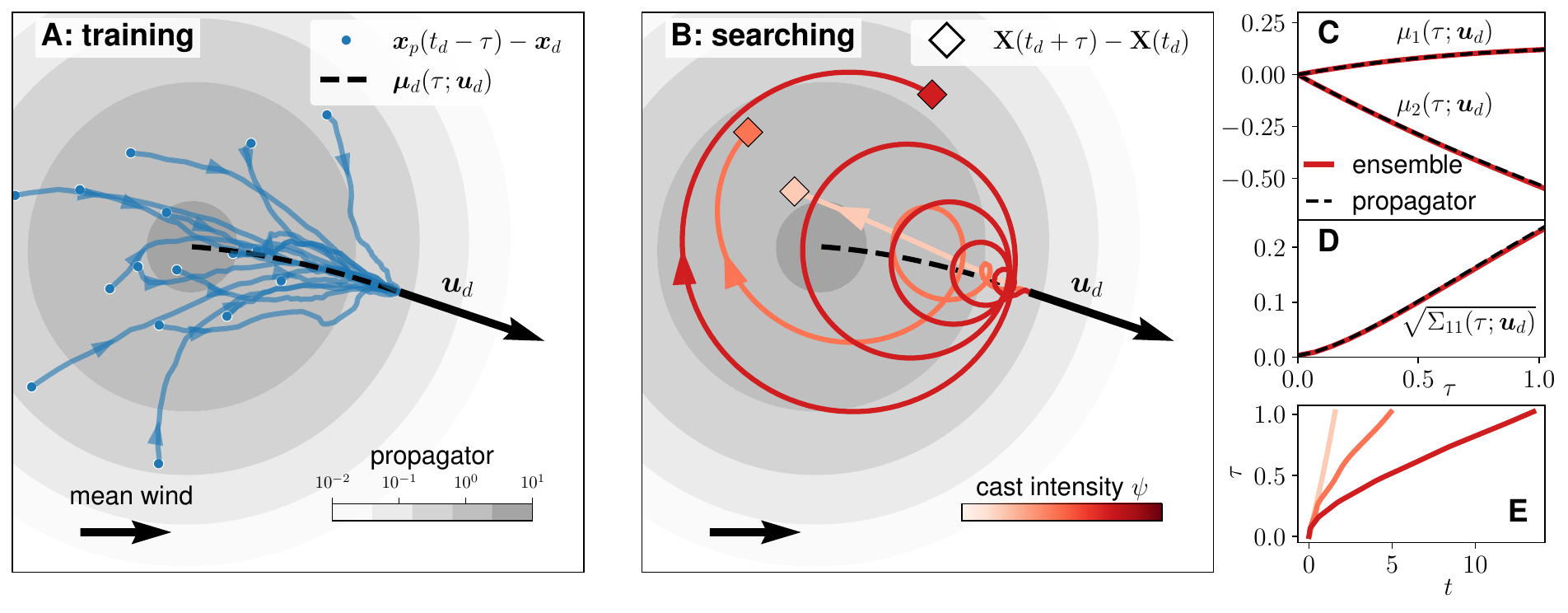}
    \vspace{-5pt}
    \caption{\textbf{Learning and sampling the backward propagator.} (a) Lagrangian trajectories of passive tracers relative to the detection point $\bm{x}_d=\bm{x}_p(t_d)$, conditioned on the local velocity $\bm{u}_d$ at detection time $t_d$. Shaded contours map the learned Gaussian backward propagator, with the dashed line tracing its evolving mean $\bm{\mu}_d(\tau;\bm{u}_d)$. 
    (b) Agent trajectories generated via the drift \eqref{eq:drift} with constant speed condition \eqref{eq:speed} for increasing casting intensity $\psi$, superimposed on the propagator.
    (c-d) Statistical validation of the inverse Fokker-Planck mapping leading to \eqref{eq:drift}, showing the components of the mean, $\mu_1(\tau)$ and $\mu_2(\tau)$, and the covariance, $\Sigma_{11}(\tau)$, obtained from the propagator against an ensemble average over agents with different initial conditions.
    (e) Time evolution of the backtracking time lag $\tau$ (see Eq.~\eqref{eq:speed}), for the trajectories shown in (b).}
    \label{fig:1}
\end{figure*}

We illustrate how source location strategies emerge from sampling a learned propagator in a statistically steady and homogeneous turbulent flow.
In an offline phase, the agent learns the backward propagator from passive tracers transported by the underlying flow,
\begin{equation}
    \frac{\de\bm{x}_p}{\de t}(t) =
        \bm{u}(\bm{x}_p,t) + \bm{U} + \sqrt{2\kappa}\,\bm{\Gamma}(t) \, ,
    \label{eq:tracers}
\end{equation}
where $\bm{x}_p$ is the tracer position, $\bm{u}$ denotes turbulent velocity fluctuations, $\bm{U}$ the mean wind, $\kappa$ the molecular diffusivity, and $\bm{\Gamma}$ a white isotropic Gaussian noise. As sketched in Fig.~\ref{fig:1}(a), for an ensemble of tracers detected at time $t_d$ and position $\bm{x}_d=\bm{x}_p(t_d)$, with local velocity fluctuation $\bm{u}_d = \bm{u}(\bm{x}_d,t_d)$, the agent estimates the conditional probability density of the detected tracer originating from $\bm{x}$ at a time $\tau$ before detection \footnote{Notice that $\tau$ does not need to coincide with physical time $t$.},
\begin{equation}
    p\left(\bm{x} + \bm{U}\tau - \bm{x}_d \middle|\tau,\bm{u}_d\right)
        = \left\langle\delta\left(
            \bm{x} - \bm{x}_p(t_d-\tau)
          \right) \middle|\tau,\bm{u}_d \right\rangle .\hspace{-2pt}
    \label{eq:propagator}
\end{equation}
Because the backward propagator $p$ describes fluctuations relative to the mean advection, it depends only on the statistics of the turbulent flow and is therefore Galilean invariant: once learned under any mean wind, it transfers to arbitrary $\bm U$ without retraining.
Moreover, in the turbulence-dominated regime, the propagator is determined primarily by flow statistics rather than by molecular diffusivity, and it can be learned from any passive tracer in the same flow, provided transport resulting from molecular diffusivity is negligible compared to turbulent transport. The tracer used for training therefore need not coincide with the chemical released by the specific source of interest during the search.

The conditional backward propagator~\eqref{eq:propagator} can be modeled using generative machine learning tools, such as normalizing flows \cite{Tabak2010,Dinh2016} or score-based models \cite{Hyvarinen2005,Song2020}. 
For simplicity and since the propagator is mainly determined by velocity, a large-scale, and often nearly-Gaussian quantity, we approximate $p$ as a Gaussian with mean $\bm\mu_d(\tau;\bm{u}_d)$ and covariance $\bm\Sigma_d(\tau;\bm{u}_d)$. Assuming statistically isotropic fluctuations, rotational invariance constrains the mean and covariance to be isotropic tensor functions of $\bm{u}_d$~\cite{Rivlin1955,Itskov2015}, namely
$\bm{\mu}_d = \alpha_d\,\bm{u}_d$ and
$\bm{\Sigma}_d = \beta_d^2\bm{I} + \left(2\beta_d\gamma_d + \gamma_d^2|\bm{u}_d|^2\right) \bm{u}_d\bm{u}_d^\top$,
ensuring that $\bm{\Sigma}_d$ is positive definite. The scalar functions $\alpha_d(\tau;|\bm{u}_d|)$, $\beta_d(\tau;|\bm{u}_d|)$, and $\gamma_d(\tau;|\bm{u}_d|)$ are parameterized through a small neural network (see End Matter for details).

Having learned the backward propagator of passive tracers, the agent has acquired a statistical model of transport in the underlying flow. This model can be used for tasks that require backward inference, such as localization of a fixed point-like source emitting a chemical. The concentration of the chemical $\theta(\bm{x},t)$ evolves as
\begin{align}
	\partial_t\theta + (\bm{u}+\bm{U})\cdot \nabla\theta = \kappa\nabla^2 \theta - \theta/\mathcal{T} + \delta(\bm{x}-\bm{Y})
    \label{eq:scal}
\end{align}
with $\kappa$ the diffusivity of the chemical, $\mathcal{T}$ a large-scale decay time, and $\bm{Y}$ the source position.
The key physical ingredient for the proposed backtracking strategy is that the concentration of a chemical at a given point is determined by the Lagrangian trajectories of passive tracers evolving backward in time toward the source~\citep{Kac1949,Risken1996}. Although the agent never observes the concentration field $\theta$ during training, tracers and chemical molecules are transported by the same flow (in a statistical sense). When the agent detects a concentration puff (high $\theta$ at the agent position), it is because the detected tracers encountered the source on their way~\cite{Falkovic2001,Celani2014}. Therefore, the agent can find the source by exploring plausible origins of the detected tracers, that is, by sampling the learned propagator~\eqref{eq:propagator}.
Furthermore, if the searching agent can remember the location $\bm{x}_d$, intensity $\theta_d=\theta(\bm{x}_d,t_d)$ and local velocity fluctuation $\bm{u}_d$ of the last $M$ detections, it can superimpose the propagators from each of the detections in memory, thus obtaining a probability map for the source position. 
Assuming conditional independence of detections, the resulting probability distribution $F(\bm{x},\tau;\bm{x}_1,\dots,\bm{x}_M;\bm{u}_1,\dots,\bm{u}_M;\theta_1,\dots,\theta_M)$ is a Gaussian with mean $\bm{\mu}(\tau)$ and covariance $\bm{\Sigma}(\tau)$,
\begin{equation}
    \ln\bprop(\bm{x},\tau) = \sum_{d=1}^M \frac{\theta_d
        \ln p\left(\bm{x}+\bm{U}\tau_d-\bm{x}_d\,\middle|\tau_d,\bm{u}_d \right) }{\sum_{l=1}^M\theta_l}
    \label{eq:belief}
\end{equation}
where $\tau_d = t_M - t_d + \tau$ is the time lag since detection at time $t_d$, $t_M$ is the time of the most recent detection, and $\tau$ is the time lag from the most recent detection~\footnote{A weighted superposition of propagators, rather than the logarithmic combination~\eqref{eq:belief}, is a viable alternative, leading to a less concentrated probability map of the source position. We use a logarithmic superposition because it preserves the Gaussian form.}. The corresponding expressions for $\bm{\mu}(\tau)$ and $\bm{\Sigma}(\tau)$ are given in the End Matter, together with the connection between \eqref{eq:belief} and the Lagrangian representation \citep{Kac1949} of the concentration field $\theta(\bm{x},t)$.

Source localization now becomes a sampling problem. Rather than designing a search strategy, we ask the inverse question: which stochastic dynamics reproduces the learned propagator, so that the probability density of the agent position coincides with $F(\bm{x},\tau)$ at every time lag $\tau$?
Let the agent follow the Langevin equation
\begin{equation}
    \frac{\de\bm{X}}{\de\tau}(\tau) = \bm{b}(\bm{X},\tau) + \sqrt{2\bm{D}(\bm{X},\tau)}\cdot\bm{\Gamma}(\tau)
    \label{eq:agent}
\end{equation}
with drift $\bm{b}$ and diffusion tensor $\bm{D}$. The probability density of $\bm{X}$ coincides with $\bprop(\bm{x},\tau)$ if it satisfies the Fokker--Planck equation associated with the agent dynamics \eqref{eq:agent},
\begin{equation}
    \partial_\tau \bprop
        = \nabla\cdot\left(-\bm{b}\bprop + \nabla\cdot(\bm{D}\bprop)\right).
    \label{eq:FP}
\end{equation}
Solving~\eqref{eq:FP} for $\bm{b}$ yields the family of drifts~\citep{Carbone2024},
\begin{equation}
    \bm{b} = \frac{\de\bm{\mu}}{\de \tau}
           + \left(\bm{\Psi}+\bm{D}-\frac{1}{2}\frac{\de \bm{\Sigma}}{\de\tau}\right)
             \cdot\nabla\ln\bprop
           + \nabla\cdot(\bm{\Psi}+\bm{D})
    \label{eq:drift}
\end{equation}
where $\bm{\Psi}^\top=-\bm{\Psi}$ is an arbitrary antisymmetric matrix function.
Equation~\eqref{eq:drift} decomposes the agent's motion into three physically distinct contributions.
First, the term $\tfrac{\de\bm{\mu}}{\de\tau}$ transports the agent along the evolving mean of the inferred source distribution. In a deterministic steady flow, it exactly retraces the detected particle's trajectory back to the source: a surge. Second, the term associated with $\bm{D}-\tfrac{1}{2}\tfrac{\de\bm{\Sigma}}{\de\tau}$ drives the agent toward regions of high probability---a chemotactic drift---while accounting for the spreading of uncertainty with increasing time lag. Finally, $\bm{\Psi}$ generates divergence-free probability currents that rotate the agent along contours of constant probability: a cast. Surge, cast, and gradient ascent thus emerge as facets of a single stochastic process rather than heuristic rules. This transport process can be interpreted as a Schr{\"o}dinger bridge~\cite{Schroedinger1931}: conditioned on detection, the learned propagator specifies the endpoint distributions linking candidate emission locations to the detection point, and the drifts~\eqref{eq:drift} generate diffusion processes whose marginals are consistent with the endpoints. Among all such processes, setting $\bm{D}=\bm{\Psi}=\bm{0}$ recovers the geodesic bridge~\cite{Bunne2023}---a direct surge.

The gauge term $\bm{\Psi}$ in~\eqref{eq:drift}, previously used to accelerate mixing in optimal transport~\cite{Villani2008} or to tune time correlations~\cite{Carbone2024}, here controls the balance between geodesic and exploratory behavior. Although the antisymmetric nature of $\bm{\Psi}$ inherently drives a casting motion, its specific functional form can be freely chosen or task-optimized, constituting a behavioral/computational choice. Here, we define it by comparing the agent's visual range $s$ with a measure of spatial uncertainty, $\sigma = [\mathrm{Tr}(\bm{\Sigma}^{-1})]^{-1/2}$ (with $\textrm{Tr}$ denoting matrix trace), which quantifies the spread of $F$. A diffuse probability map ($\sigma \gg s$) demands exploration along the probability contours, whereas a peaked $F$ ($\sigma \ll s$) should collapse the trajectory into a geodesic surge. In two dimensions, a minimal choice that smoothly bridges these limits is $\Psi_{12} = \psi \,\mathcal{U} \, \sigma/(s + \sigma)$ with $\mathcal{U}$ a reference speed and $\psi$ the dimensionless casting intensity.
Trajectories for various $\psi$ are shown in Fig.~\ref{fig:1}(b), illustrating the transition from geodesic paths to spiral exploration of probability contours.

So far, the coupling between the physical time $t$ and the backtracking time lag $\tau$---which dictates how far into the past the agent is sampling---is not specified, representing a degree of freedom of the backtracking strategy. For comparison with classical heuristics, we consider deterministic paths ($\bm{D}=\bm{0}$) traversed at a constant speed, $\vert\de\bm{X}/\de t\vert = \mathcal{U}$, requiring
\begin{equation}
    \frac{\de\tau}{\de t}(t) = \frac{\mathcal{U}}{|\bm{b}(\bm{X}(\tau(t)),\tau(t))|}.
    \label{eq:speed}
\end{equation}
At the moment of detection, $\tau=0$ and $\bm{X}(0)=\bm{x}_M$ (the position of the most recent detection), $F$ is nearly singular and is regularized by the finite agent size $a$, namely $\bm{\Sigma}(0) = a\bm{I}$, with $\bm{I}$ the identity matrix. The mean of $F$ is initialized as $\bm{\mu}(0)=\bm{x}_M+\bm{\epsilon}$, where $\bm{\epsilon}$ is a Gaussian random displacement $\bm{\epsilon}\sim \mathcal{N}(\bm{0},a^2\bm{I})$. Randomness in the initial condition for $\bm{\mu}$ serves to push the agent trajectory away from the propagator mean when $\bm{D}=\bm{0}$. Figures~\ref{fig:1}(c-d) show that the mean and covariance of the agent position computed from an ensemble of agents coincide with the mean and covariance of the learned propagator. Although this holds for all casting intensities, a larger $\psi$ corresponds to a finer sampling of the propagator, and thus a slower evolution of the backtracking time lag $\tau$ as a function of time $t$, as shown in Fig.~\ref{fig:1}(e).

We quantify the performance of the backtracking strategy for source localization in a concentration field advected by a statistically steady, homogeneous, incompressible two-dimensional turbulent flow with an inverse energy cascade~\cite{Boffetta2012} under three mean-wind regimes: isotropic ($\bm{U}=\bm{0}$), moderate ($|\bm{U}|=u'$), and strong ($|\bm{U}|=2u'$), where $u'$ is the root-mean-square velocity. 
The agents sample the local concentration and velocity every $\Delta t = 2a/\mathcal{U}$. Detection occurs whenever the concentration at the agent position exceeds a threshold $\theta^*$. We initialize the agents at random locations where $\theta\ge\theta^*$ and let them evolve by integrating Eqs.~(\ref{eq:agent},\ref{eq:speed}) with time step $\Delta t$. An agent reaches the source when their distance is below the visual range $|\bm{X}-\bm{Y}|\le s$, while trajectories longer than a maximum length are classified as lost. For details on the numerical setup, see End Matter.

\begin{figure}
    \centering
    \includegraphics[width=\columnwidth]{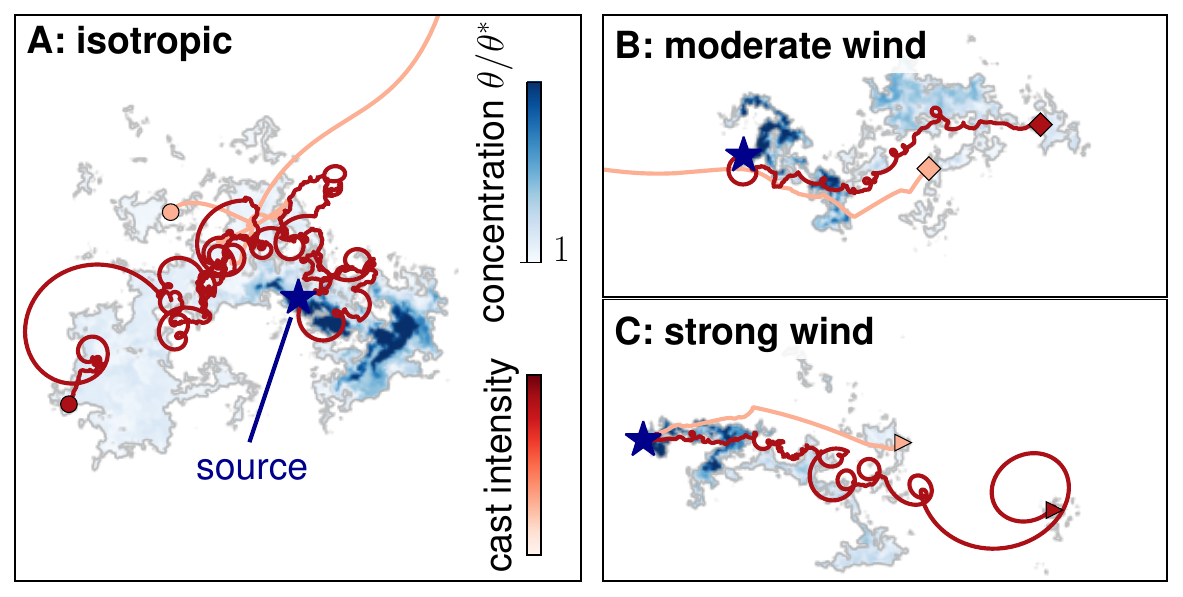}
    \caption{\textbf{Search trajectories in two-dimensional turbulence.} Agent paths driven by the backtracking strategy at varying casting intensities $\psi$ and different mean winds. The background displays the initial underlying concentration resolved above the agent's sensitivity threshold $\theta^*$. Only subsets of the full simulation domain are shown.}
    \label{fig:2}
\end{figure}
\begin{figure}
    \centering
    \includegraphics[width=\columnwidth]{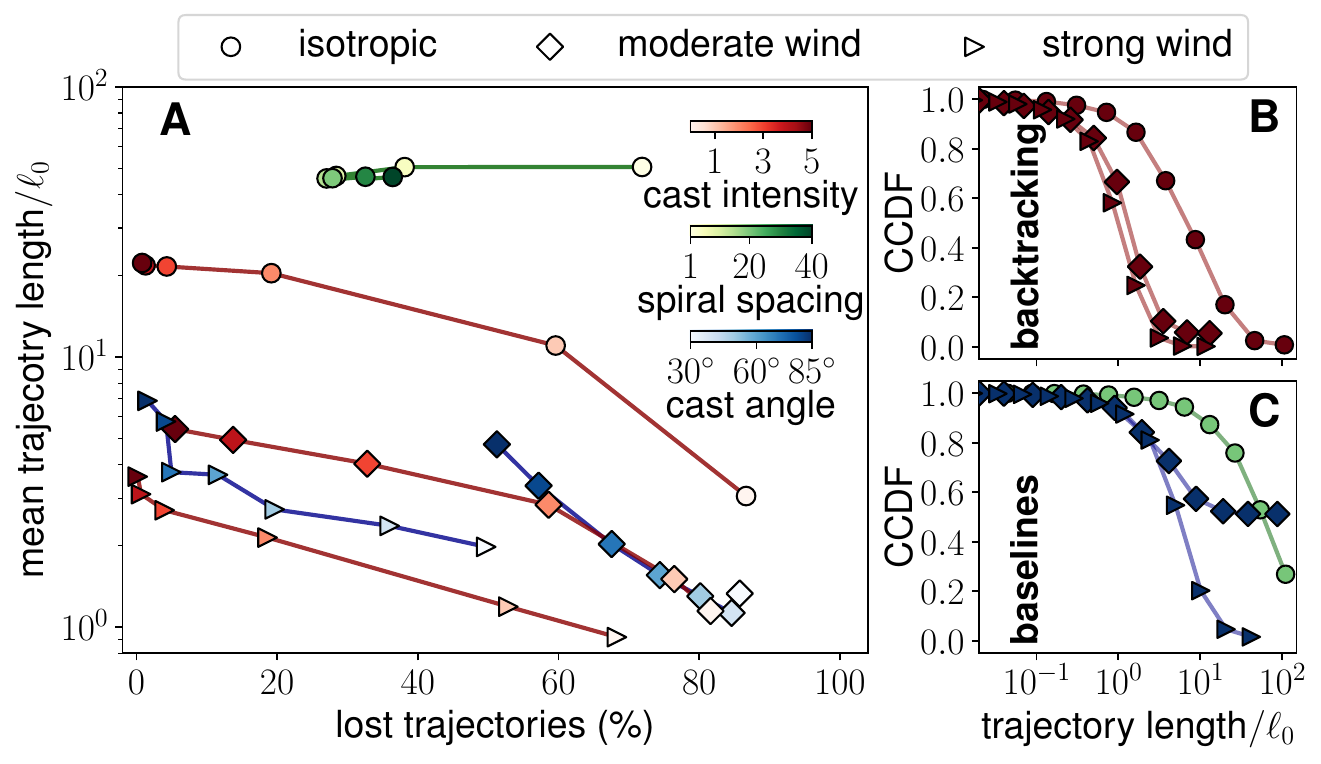}
    \caption{\textbf{Performance of the backtracking strategy against classical baselines.} (a) Mean trajectory length, conditioned on finding the source, versus the fraction of lost trajectories for the backtracking (red) at various casting intensities $\psi$, and for the baseline heuristics (green, blue). The results are for three mean wind regimes: isotropic ($\bm{U}=\bm{0}$), moderate ($|\bm{U}|=u'$), and strong ($|\bm{U}|=2u'$). Heuristic baselines consist of an Archimedean spiral with varying branch spacing (expressed in agent size units, $a$) for $\bm{U}=\bm{0}$, and cast-and-surge with varying cone apertures for $|\bm{U}| > 0$. (b) Complementary cumulative distribution functions (CCDF) of the trajectory lengths for backtracking and (c) baselines for the parameters yielding minimum fraction of lost trajectories. 
    Statistics are computed over an ensemble of agents starting at random positions where $\theta\ge\theta^*$ and $|\bm{x}-\bm{Y}|>s$.
    The reference length is the agents' initial average distance from the source $\ell_0=\langle|\bm{X}(0)-\bm{Y}|\rangle$ (see End Matter for its values).}
\label{fig:3}
\end{figure}

Representative trajectories are shown in Fig.~\ref{fig:2}: as the casting intensity $\psi$ increases, the nearly geodesic backtracking progressively turns into an exploratory sampling of probability contours, trading longer trajectories for a marked reduction in the fraction of lost searches. Across all wind regimes, the family of backtracking strategies obtained by varying $\psi$ generates a continuous trade-off in the plane of mean trajectory length versus lost fraction (see red curves in Fig.~\ref{fig:3}(a)).

We compare the backtracking strategy against two classical baselines: cast-and-surge~\cite{Balkovsky2002}, optimizing the surge duration and cone aperture, and Archimedean spiral search~\cite{Masson2009} in the isotropic case, optimizing the spiral branch spacing. For cast-and-surge, we report the tuning of the surge steps that minimizes the fraction of lost trajectories at each casting angle. Figure~\ref{fig:3}(a) shows that, in all flow regimes, backtracking achieves a shorter mean trajectory (conditioned on finding the source) than the best-tuned baseline at fixed lost fraction. In particular, by increasing $\psi$, the lost fraction can be essentially driven to zero while incurring a modest penalty in arrival time.
The complementary cumulative distributions of trajectory length in Fig.~\ref{fig:3}(b,c) indicate that the improvement is not limited to average performance: backtracking also suppresses the heavy tails associated with extremely long trajectories. All results are obtained from a single propagator trained in zero-wind conditions and transferred to the wind regimes via Galilean invariance, without retraining, a degree of transferability that is unavailable to strategies derived from repeated search trials in a given concentration field.

\begin{figure}
    \centering
    \includegraphics[width=\columnwidth]{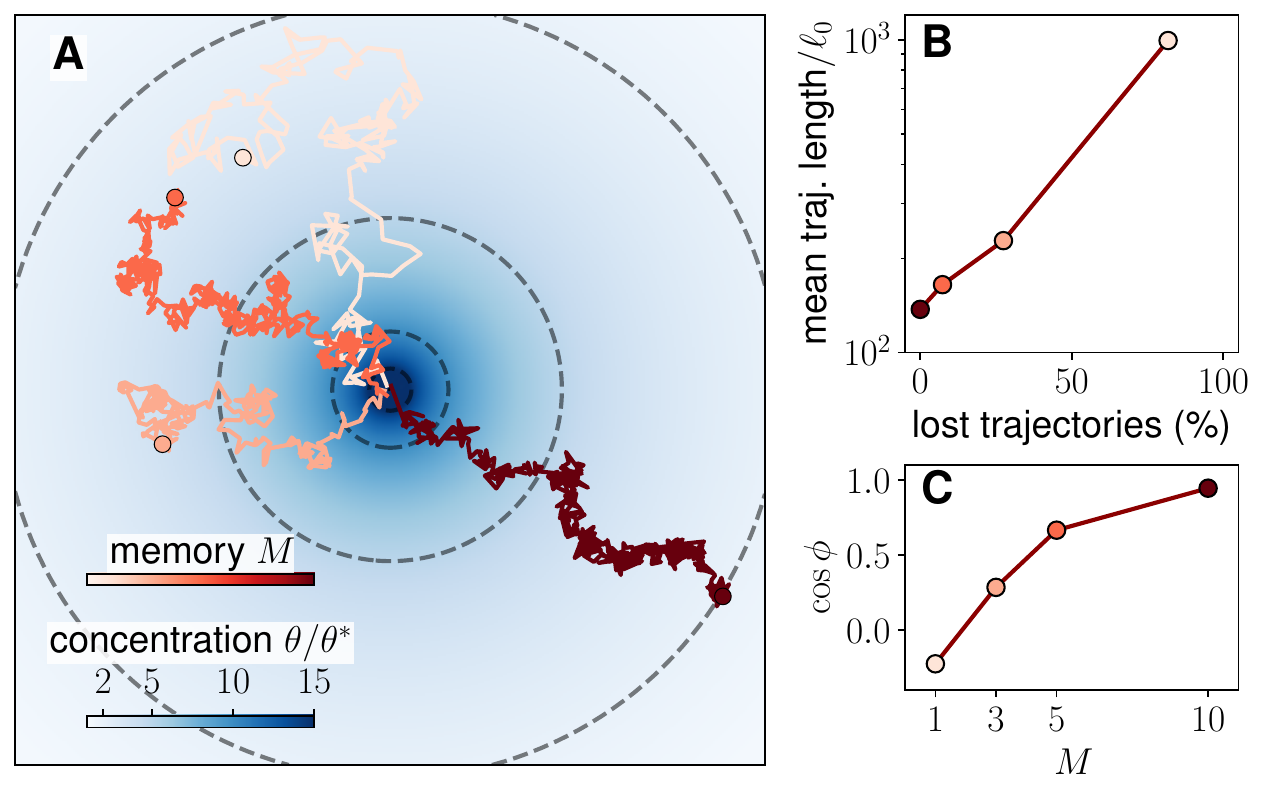}
    \caption{\textbf{Chemotactic limit.} (a) Paths of agents retaining the latest $M$ detections in memory. (b) Mean trajectory length normalized with the initial average distance of the agents from the source versus fraction of lost trajectories.
    (c) Statistical alignment between the time-averaged drift and concentration gradient at the agent position, $\cos\phi=\langle\overline{\bm{b}}\cdot\overline{\nabla\theta}/(|\overline{\bm{b}}|\,|\overline{\nabla\theta}|)\rangle$, where the overbar denotes time average.}
\label{fig:chemotaxis}
\end{figure}

The results presented for source localization in two-dimensional turbulence are for agents retaining in memory only the most recent detection ($M=1$). This suffices for robust performance because the turbulent flow provides directional information for backtracking. We now consider the opposite physical limit: the stringent case in which the flow is removed altogether ($\bm{U}=\bm{u}=\bm{0}$). In this pure diffusion limit, the agent must rely entirely on its memory of past detections to infer the chemical landscape~\cite{Celani2010}. We ask whether the drift \eqref{eq:drift}, derived from backward transport, naturally reduces to a chemotactic dynamics that guides the agent up the concentration gradient.
In this regime, the backward propagator for a single detection, regularized with the agent's finite size, is a zero-mean Gaussian with covariance $(2\kappa\tau_d+a^2) \bm{I}$. Setting $\bm{\Psi}=\bm{0}$ (not essential for gradient ascent), and for $\bm{D}$ independent of $\bm{x}$, the drift \eqref{eq:drift} features the score $\nabla\ln F$ and the transport of the mean, $\de\bm{\mu}/\de\tau$. For an agent aggregating $M$ detections at positions $\bm{x}_d$--independently distributed with centroid $\bm{x}_0$ and spatial covariance $\bm{R}$--we Taylor expand the local concentration field about $\bm{x}_0$. To leading order in $|\bm{R}|$, the score and transport terms projected along the concentration gradient at $\bm{x}_0$ are
\begin{subequations}
\begin{align}
    \left\langle\nabla\ln F \cdot \nabla\ln\theta_0 \right\rangle &\sim \frac{M-1}{M} S_1 |\nabla\ln\theta_0|^2_{\bm{R}} \label{eq:chemotaxis_score}
\\
    \left\langle \frac{\de\bm{\mu}}{\de\tau} \cdot \nabla\ln\theta_0 \right\rangle &\sim \frac{4\kappa}{M} \frac{S_1 S_3 - S_2^2}{S_1^3} |\nabla\ln\theta_0|^2_{\bm{R}}
\label{eq:chemotaxis_transp}
\end{align}
\label{eq:chemotaxis}
\end{subequations}
with $S_n = \sum_{d=1}^M (2\kappa\tau_d+a^2)^{-n}/M$, and vector norm $|\bm{v}|^2_{\bm{R}}=(\bm{v} \cdot \bm{R})\cdot\bm{v}$. 
Both terms in \eqref{eq:chemotaxis} are positive for $M>1$, implying statistical alignment between the score and transport terms in drift \eqref{eq:drift} with the concentration gradient (see End Matter for the derivation of \eqref{eq:chemotaxis}).
We verify the analytical prediction \eqref{eq:chemotaxis} by simulating source localization in a smooth isotropic concentration field with an ensemble of agents that retain the latest $M$ detections in memory. We integrate \eqref{eq:agent}, setting $\tau=t$, $\bm{\Psi}=\bm{0}$, and $\bm{D}=\epsilon\bm{I}$, where the constant diffusivity $\epsilon$ controls the intensity of noise (see Fig.~\ref{fig:chemotaxis}(a) and End Matter). For $M=1$, agents that reach the source do so primarily by random exploration. As $M$ increases, the fraction of lost trajectories and the mean trajectory length decrease, as shown in Fig.~\ref{fig:chemotaxis}(b). This improvement is driven by the alignment between the trajectories and the concentration gradient averaged along the agent's path (see Fig.~\ref{fig:chemotaxis}(c)), indicating the emergence of chemotactic behavior.

In summary, learning the underlying transport provides a principled route from physical dynamics to search strategies.
By acting on the propagator of the tracers instead of directly on the concentration field, exploiting the Lagrangian representation of scalar transport, the proposed framework achieves interpretability, transferability, and efficiency across heterogeneous flow conditions. The central result is the family of drifts in Eq.~\eqref{eq:drift}, obtained by solving an inverse Fokker--Planck problem for a superposition of backward propagators of passive tracers. Within this unified formulation, classical chemotactic and cast-and-surge strategies naturally emerge as complementary aspects of a single transport-based principle, rather than ad hoc behavioral rules.

This perspective opens several directions for systematic extensions, including task-dependent optimization of exploratory components encoded in $\bm{\Psi}$ and $\bm{D}$, and applications to the localization of moving sources~\cite{Carbone2026}. More broadly, the proposed backtracking framework applies to any setting in which transport statistics can be learned offline and subsequently used to locate sources from local observations. Beyond physical flows, analogous inverse problems arise in diffusion and spreading processes in complex networks~\cite{Satorras2015} and generative models~\cite{Song2020,DeBortoli2021}.

\begin{acknowledgments}
We acknowledge fruitful discussions with Massimo Cencini, Luca Biferale, Michele Buzzicotti, Antonio Celani, Robin Heinonen, and Agnese Seminara.
We acknowledge financial support under the National Recovery and Resilience Plan (NRRP), Mission 4, Component 2, Investment 1.1, Call for tender No.~104 published on 2.2.2022 by the Italian Ministry of University and Research (MUR), funded by the European Union--NextGenerationEU--Project Title Equations informed and data-driven approaches for collective optimal search in complex flows (CO-SEARCH), Contract 202249Z89M--CUP B53-D23003920006 and E53-D23001610006.
This work was also supported by the Italian Ministry of University and Research (MUR)--Fondo Italiano per la Scienza (FIS2)--2023 Call, project DeepFL, CUP E53C24003760001, and by the European Research Council (ERC) under the European Union's Horizon 2020 research and innovation program (Grant Agreement No.~882340).
\end{acknowledgments}

\bibliography{apssamp}

\appendix

\setcounter{figure}{0}
\renewcommand\thefigure{A\arabic{figure}}
\setcounter{equation}{0}
\renewcommand\theequation{A\arabic{equation}}

\section{End Matter}

\subsection{Details on the numerical simulations}

The statistically steady incompressible two-dimensional turbulent flow used to learn the propagator and for source localization tests is resolved on a $N^2=1024^2$ periodic grid with spacing $\Delta x=2\pi/N$, time step $\Delta t_{\rm DNS}=2\times10^{-4}$, root-mean-square velocity $u'=0.4$, and small-scale time $\tau_\Omega=\langle|\nabla\times\bm{u}|^2\rangle^{-1/2}\simeq0.05$. Turbulence is maintained by Gaussian forcing with characteristic scale $0.03$, combined with large-scale damping, yielding an integral scale of order one. The diffusivity of the scalar is $\kappa=2\times10^{-4}$, the decay time scale is $\mathcal{T}=20$, and the scalar is absorbed at the domain boundaries to avoid periodicity effects.
Additional details on the numerical simulations are given in \cite{Piro2026}.

A dataset of $10^5$ passive tracer trajectories from these simulations, sampled every $\Delta t=160\,\Delta t_{\rm DNS}$, is used to learn the backward propagator \eqref{eq:propagator} in an offline phase. The time-lag-dependent mean and covariance of the Gaussian model are represented by a multilayer perceptron with two inputs (time lag and flow speed at detection), three hidden layers of 16 neurons, and three outputs corresponding to the isotropic tensor functions $\alpha_d$, $\beta_d$, and $\gamma_d$. Training is implemented in PyTorch \citep{PyTorch2019} by maximum-likelihood estimation of the prescribed Gaussian model \eqref{eq:propagator} on the tracer dataset \citep[e.g.][]{Husmeier2013}.

The learned propagator \eqref{eq:propagator} provides a reduced model of backward transport, connecting the tracer dynamics \eqref{eq:tracers} to the concentration field \eqref{eq:scal} through the Feynman--Kac representation \citep{Kac1949,Risken1996}
\begin{align}
\theta(\bm{x},t;\bm{Y})
=
\int_0^\infty \de\tau
e^{-\tau/\mathcal{T}}
\,G\!\left[\bm{Y},t-\tau\middle|\bm{x},t;\bm{u}+\bm{U}\right]
\label{eq:FeynmanKac}
\end{align}
where $G$ is the exact backward propagator. Inferring the source position $\bm{Y}$ from local concentration measurements amounts to inverting \eqref{eq:FeynmanKac}. However, the exact propagator $G$ is a functional of the entire space--time velocity field realization and is inaccessible to an agent that relies on local detections. The learned propagator \eqref{eq:propagator} replaces this intractable object with a single-point conditional approximation. This approximation reflects the limited sensing capabilities of the agent and the need for a tractable representation of backward transport. We also assume that the scalar lifetime $\mathcal{T}$ is much longer than the characteristic time between detections, neglecting the exponential decay of remote emissions.

Once learned, the propagator \eqref{eq:propagator} is used for source localization in two-dimensional turbulence, under different mean-wind conditions without retraining. The agents have size $a=\Delta x$ and sample the local concentration and velocity every $\Delta t = 2a/\mathcal{U}$, where $\mathcal{U}=\max(|\bm{U}|,u')$ is the agent speed. We initialize $10^4$ agents at random locations where $\theta(\bm{x},t)\ge\theta^*$ and $|\bm{x}-\bm{Y}|>s$. We set the sensitivity threshold to $\theta^*=1.33$, the visual range to $s=10a$, and define as lost the trajectories of length exceeding $2\cdot10^4\,a$. The mean initial distance of the agents from the source, $\ell_0=\langle|\bm{X}(0)-\bm{Y}|\rangle$, is approximately 180, 200, and 236 agent sizes $a$ in the zero-, moderate-, and strong-wind cases, respectively. It is representative of the spatial extent of the concentration field and is used as reference length in Fig.~\ref{fig:3}.

Finally, for chemotactic tests, we consider the limit of pure diffusion, opposite to turbulence-dominated transport. For $M>1$, the agent uses the Gaussian propagator to construct the probability map \eqref{eq:belief}, that is, a Gaussian with $\tau$-dependent covariance and mean
\begin{align}
\bm{\Sigma}^{-1} = \sum_{d=1}^M \tilde{\theta}_d \bm{\Sigma}_d^{-1},  &&
\bm{\mu} = \bm{\Sigma}\cdot \sum_{d=1}^M \tilde{\theta}_d \bm{\Sigma}_d^{-1}\cdot\tilde{\bm{\mu}}_d,
\label{eq:param_belief}
\end{align}
where $\tilde{\theta}_d = \theta_d/\sum_{d'=1}^M\theta_{d'}$ is the normalized detection intensity and $\tilde{\bm{\mu}}_d=\bm{\mu}_d+\bm{x}_d-\bm{U}\tau_d$ is the propagator mean shifted to the detection location and by the displacement induced by the mean wind.
The concentration field is the steady solution of \eqref{eq:scal} with $\bm{u}=\bm{U}=\bm{0}$, $\mathcal{T}=10$ and $\kappa=0.1$, namely $\theta(\bm{x})\propto K_0 \left(|\bm{x}-\bm{Y}|\right)$, where $K_0$ is the modified Bessel function of the second kind. In this limit, the Gaussian propagator supplied to the agent is exact and coincides with $G$ in \eqref{eq:FeynmanKac}. The agent diffusivity is set to $\epsilon=10^{-3}$, while the spatial and temporal resolutions are unchanged from the zero-mean-wind turbulent simulations. An ensemble of $10^4$ agents is initialized uniformly at points where $|\bm{x}-\bm{Y}|>s$, with a visual range $s=10a$, giving an average initial distance of the agents from the source $\ell_0=352\,a$ used as reference length in Fig.~\ref{fig:chemotaxis}. The agent speed is not prescribed and, since the trajectories are stochastic, the mean trajectory length reported in Fig.~\ref{fig:chemotaxis}(b) scales with the simulation time step as $1/\sqrt{\Delta t}$.

\subsection{Derivation of Equation \eqref{eq:chemotaxis}}

To derive the contributions to the effective drift \eqref{eq:drift} in the pure-diffusion limit, we consider an agent sampling at positions $\bm{x}_d$ at time lags $\tau_d$ from the current time, retaining in memory the latest $M$ detections. The sampling is discrete with a time interval $\Delta t$. Assuming a smooth field, we Taylor-expand the concentration $\theta(\bm{x})  \sim \theta_0 + \nabla\theta_0 \cdot \bm{r}_d$. Here, $\bm{x}_0 = \sum_d \bm{x}_d/M$ is the centroid of the agent's positions, $\theta_0 = \theta(\bm{x}_0)$ is the concentration at the centroid, and $\bm{r}_d = \bm{x}_d - \bm{x}_0$ are the relative displacements, small with respect to the typical length scale of $\theta(\bm{x})$.
If the sampling positions are independent, with single-time spatial covariance $\bm{R}$, the relative displacements have covariance $\langle\bm{r}_d\bm{r}_{d'}^T\rangle = \bm{R}(\delta_{dd'} - 1/M)$. The regularized backward propagator has spatial covariance $(2\kappa\tau_d + a^2)\bm{I}$, where $\kappa$ is the diffusivity of the chemical and $a$ is the characteristic size of the agent. 

The score associated with the probability map \eqref{eq:belief} evaluated in the centroid $\bm{x}_0$ is $\nabla\ln F = \bm{n}/\Theta$ with $\Theta=\sum_d\theta_d$, $\bm{n} = \sum_d \theta_d w_d \bm{r}_d$, and $w_d = (2\kappa\tau_d + a^2)^{-1}$. This is obtained through \eqref{eq:param_belief} with $\bm{U}=\bm{0}$ and $\bm{\mu}_d=\bm{0}$ in the purely diffusive regime.
Because $\sum_d \bm{r}_d = \bm{0}$, normalization reduces to $\Theta = M\theta_0$ to first order in $|\bm{r}_d|$. 
Expanding the numerator to first order in $|\bm{r}_d|$ yields
\begin{align}
\bm{n} \sim \theta_0\sum_d w_d \bm{r}_d + \sum_d w_d(\nabla\theta_0\cdot\bm{r}_d)\bm{r}_d.
\end{align}
Averaging over the agent's path realizations and using the covariance of the displacements results in the score
\begin{align}
\langle \nabla\ln F \rangle \sim
\frac{M-1}{M} S_1 \bm{R} \cdot \nabla\ln\theta_0
\end{align}
with $S_n=\sum_d w_d^n / M$. Taking the dot product with the log-derivative of the concentration field at the centroid yields the gradient alignment \eqref{eq:chemotaxis_score}.

An analogous expansion and averaging gives the asymptotic mean of the probability map over the inferred source position \eqref{eq:belief}, $\bm{\mu} = \bm{x}_0 + \bm{n}/(\sum_d \theta_d w_d)$. Expanding to first order in $|\bm{r}_d|$ and averaging using the covariance of the agent's displacements yields
\begin{align}
\langle\bm\mu\rangle \sim \bm{x}_0 + \left( 1-\frac{S_2}{M S_1^2} \right) \bm{R}\cdot \nabla\ln\theta_0.
\end{align}
Differentiating the mean with respect to $\tau$ and using $\de S_n/\de\tau = -2n\kappa S_{n+1}$, we finally get
\begin{align}
\left\langle\frac{\de\bm{\mu}}{\de\tau}\right\rangle \sim \frac{4\kappa}{M}\frac{S_1S_3-S_2^2}{S_1^3} \bm{R}\cdot  \nabla\ln\theta_0.
\end{align}
The factor $S_1S_3-S_2^2$ is positive by the Cauchy-Schwarz inequality, namely $(\sum w_d)(\sum w_d^3) \geq (\sum (w_d^{1/2} w_d^{3/2}))^2$. This leads to statistical alignment between the transport term in the agent's drift \eqref{eq:drift} and the concentration gradient at the centroid of the last $M$ agent's positions \eqref{eq:chemotaxis_transp}.  Furthermore, evaluating the score and transport terms at the centroid $\bm{x}_0$ rather than at the current position of the agent $\bm{x}_M$ leaves the prediction \eqref{eq:chemotaxis} unchanged to the leading order in $|\bm{r}_d|$.
Equation \eqref{eq:chemotaxis} represents a conservative estimate of statistical alignment, since it assumes that the agent's previous displacements are independent and uncorrelated with the underlying concentration field. This models a ``worst-case'' scenario of random exploration. Despite this random walk setup, the backtracking strategy naturally induces chemotactic behavior.

\end{document}